\documentclass[reprint,twocolumn,aps,superscriptaddress,showpacs]{revtex4}
\usepackage{amssymb}
\usepackage{amsmath}
\usepackage{graphicx}
\usepackage[normalem]{ulem}
\usepackage[dvips]{color}
\usepackage{graphicx}
\usepackage{subfigure}

\renewcommand\sout{\bgroup \color{red} \ULdepth=-.5ex \ULset}

\begin{document}

\title{The influence of the symmetry energy on the cone-azimuthal emission}
\author{Yuan Gao}
\email{gaoyuan@impcas.ac.cn}

\affiliation{School of Information Engineering, Hangzhou Dianzi
University, Hangzhou 310018, China}

\author{G.C. Yong}
\affiliation{Institute of Modern Physics, Chinese Academy of
Sciences, Lanzhou 730000, China}

\author{Yongjia Wang}
\affiliation{School of Science, Huzhou Teachers College, Huzhou
313000, China}
\affiliation{School of Nuclear
Science and Technology, Lanzhou University, Lanzhou 730000, China}

\author{Qingfeng Li}
\affiliation{School of Science, Huzhou Teachers College, Huzhou
313000, China}

\author{Wei Zuo}
\affiliation{Institute of Modern Physics, Chinese Academy of
Sciences, Lanzhou 730000, China}

\begin{abstract}

In the framework of the isospin-dependent
Boltzmann-Uehling-Uhlenbeck transport model, effects of the
symmetry energy on the evolutions of free n/p ratio and charged
pion ratio in the semi-central collision of $^{197}$Au+$^{197}$Au
at an incident beam energy of 400 MeV/nucleon are studied. At the
beginning of the reaction (before 11 fm/c) they are both affected
by the low-density behavior of the symmetry energy but soon after
are affected by the high-density behavior of the symmetry energy
after nuclei are compressed (after 11 fm/c) and the effects of the
symmetry energy are generally smaller compared with the central
collision case. Interestingly, their dependences on the symmetry
energy are shown to arise with increase of cone-azimuthal angle of
the emitted particles. In the direction perpendicular to the
reaction plane, the $\pi^{-}/\pi ^{+}$ ratio or free n/p ratio
especially at high kinetic energies exhibits significant
sensitivity to the symmetry energy.

\end{abstract}

 \maketitle

%\section{Introduction}
One of the main topics in the nuclear physics is to give a
detailed description of the equation of state (EoS) of asymmetric
nuclear matter. The EoS at density $\rho$ and isospin asymmetry
$\delta$ ($\delta=(\rho_n-\rho_p)/(\rho_n+\rho_p)$) is usually
expressed as \cite{li08,bar05}
\begin{equation}
E(\rho ,\delta )=E(\rho ,0)+E_{\text{sym}}(\rho )\delta ^{2}+\mathcal{O}%
(\delta ^{4}),
\end{equation}%
where $E_{\text{sym}}(\rho)$ is nuclear asymmetry energy. Around
normal nuclear matter density, the symmetry energy has a value of
about 30 MeV, e.g., from fitting the binding energies of atomic
nuclei with the liquid- drop mass formula, but its behavior at
supranormal densities is still poor known. And yet it is very
important for understanding the structure and evolution of many
astrophysical objects such as neutron stars, supernova explosions,
etc. In general, values of the symmetry energy at different
densities can be obtained by many-body microscopic theories.
However, predictions from various many-body approaches diverge
widely, even with the same model but different parameters. New
light has been thrown on this puzzle with the advent of modern
accelerators and suitable detectors. The planning radioactive beam
facilities at CSR (China), FAIR (Germany), RIKEN (Japan),
SPIRAL2/GANIL (France) and the upcoming FRIB (USA) provide an
unique opportunity to study experimentally the EoS of asymmetric
nuclear matter. To extract information from experimental data, it
is challenging to find out observables which are sensitive enough
to the symmetry energy. Much effort has been devoted during the
last decade\cite{li08,bar05,trautmann12}. A number of observables
are proposed, such as the free neutron/proton ratio \cite{li04b},
the isospin fractionation\cite{Muller95,Baran02}, the
neutron-proton correlation function\cite{chen03a},
t/$^{3}$He\cite{chen03b}, the isospin diffusion \cite{Shi03}, the
neutron-proton transverse differential flow \cite{Scalone99}, the
elliptic collective flows\cite{li01,liqf11,cozma11}, the
$K^{0}/K^{+}$ ratio\cite{fer06} and the $\pi^{-}/\pi ^{+}$
ratio\cite{li02,Gai04}, etc. Unfortunately, the information
extracted from experimental data is so confused. From the FOPI
data\cite{rei07}, Some predicted a stiff
dependence\cite{feng10,russotto11}, whereas other conclusions
imply the opposite situation, i.e the symmetry energy would
decrease at certain suprasaturation densities \cite{xiao09,xie13}.
While studying model dependence of isospin sensitive observables
of nuclear symmetry energy \cite{gwm13}, a model-independent study
of the high-density dependence of the symmetry energy has been
carried out by Cozma, \emph{et al}. \cite{cozma2013}, which
constrain nuclear symmetry energy more \emph{scrupulously}. To
further constrain the EoS of asymmetric nuclear matter, it is
worthwhile finding out more delicate observables to probe the high
density behavior of the nuclear symmetry energy.

The anisotropic distribution of particle emissions in c.m. system
has been studied for quite some time and it has been well known
that in the direction perpendicular to the reaction plane the
particles emitted carry more information of the squeezed-out dense
nuclear matter \cite{yongso,msup}. Here we revisit this question,
as it may be an effective tool to probe the symmetry energy at
high densities.

%\section{The theoretical model}
Studying isospin physics in the heavy-ion collision at
intermediate energies, the isospin-dependent
Boltzmann-Uehling-Uhlenbeck transport model (IBUU) has been a
successful approach to describing the dynamical evolution of the
systems in phase space. In this model, the mean-field potential
(MDI) can be written as\cite{Das03}
\begin{eqnarray}
U(\rho, \delta, \textbf{p},\tau)
=A_u(x)\frac{\rho_{\tau^\prime}}{\rho_0}+A_l(x)\frac{\rho_{\tau}}{\rho_0}\nonumber\\
+B\left(\frac{\rho}{\rho_0}\right)^\sigma\left(1-x\delta^2\right)\nonumber
-8x\tau\frac{B}{\sigma+1}\frac{\rho^{\sigma-1}}{\rho_0^\sigma}\delta\rho_{\tau^{\prime}}\nonumber\\
+\sum_{t=\tau,\tau^{\prime}}\frac{2C_{\tau,t}}{\rho_0}\int{d^3\textbf{p}^{\prime}\frac{f_{t}(\textbf{r},
\textbf{p}^{\prime})}{1+\left(\textbf{p}-
\textbf{p}^{\prime}\right)^2/\Lambda^2}},
\label{Un}
\end{eqnarray}
where $\tau=1/2$ ($-1/2$) for neutrons (protons) and the
$f_{t}(\textbf{r}, \textbf{p})$ is the phase space distribution
function which is solved following a test particle evolution on a
lattice. The $\delta=(\rho_n-\rho_p)/(\rho_n+\rho_p)$ denotes the
isospin asymmetry of nuclear medium. By varying the variable $x$
this potential can be used to get different forms of the symmetry
energy predicted by various many-body theories without changing
any property of symmetric nuclear matter and the value of symmetry
energy at normal density $\rho_0$. In this work we choose the
situations $x=0$ (stiff) and $x=1$ (soft) as comparison. The
quantities $A_{u}(x),A_{l}(x)$ are $
A_{u}(x)=-95.98-\frac{2B}{\sigma +1}x $, $
A_{l}(x)=-120.57+\frac{2B}{%
\sigma +1}x $, here $B=106.35$ MeV. $\Lambda =p_{F}^{0}$ is the
nucleon Fermi momentum in
symmetric nuclear matter, $C_{\tau ,\tau ^{\prime }}=-103.4$ MeV and $%
C_{\tau ,\tau }=-11.7$ MeV. The $C_{\tau ,\tau ^{\prime }}$ and
$C_{\tau ,\tau }$ terms are the momentum-dependent interactions of
a nucleon with unlike and like nucleons in the surrounding nuclear
matter. The corresponding incompressibility of symmetric nuclear
matter at saturation density is $211$ MeV. Here the symmetry
potential is momentum dependent. Its effects on the free $n/p$
ratio, the $\pi$ production and other observables
\cite{li04a,gao11,zhang12,gao12} were investigated. Another
important ingredient in heavy-ion collisions is the
nucleon-nucleon (NN) cross sections. Medium effects on the NN
elastic cross sections have not been well determined so far. In
our calculations, the reduction scale according to nucleon
effective masses is adopted \cite{li05a}. This modification is
isospin- and momentum-dependent and similar to the BHF with
three-body force or DBHF approach calculations only if nucleonic
momentum is not too large \cite{ligq93,fuchs01,zhf07,zhf10}. The
total and differential cross sections for all other particles are
taken either from experimental data or obtained by using the
detailed balance formula.

In the following, to see the correlation between the anisotropic
distribution of the azimuthal angle of the particle emission and
effects of the symmetry energy in heavy-ion collisions more
specifically \cite{yongso,msup}, we use 40,0000 semi-central
$^{197}$Au+$^{197}$Au reaction events for each cases to study the
influence of the symmetry energy on the azimuthal emission.

%\section{Results and discussions}

\begin{figure} \centering
\subfigure[~Effect of the symmetry energy on the evolution of free
n/p ratio.] { \label{ek-np}
\includegraphics[height=5.5cm,width=0.8\columnwidth]{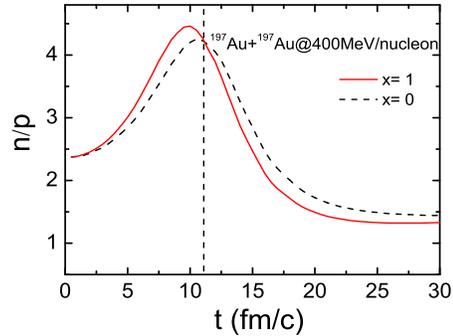}
} \subfigure[~Effect of the symmetry energy on the evolution of
charged pion ratio.] { \label{ek-pi}
\includegraphics[height=5.5cm,width=0.8\columnwidth]{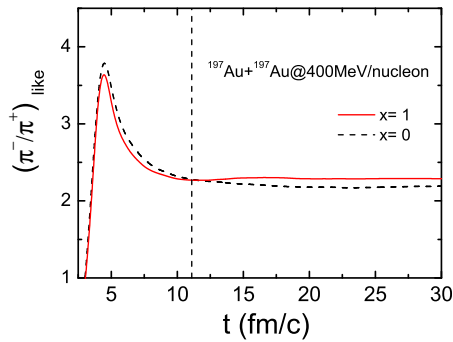}
} \caption{(Color online) Evolutions of the particle ratios with different
symmetry energies in the reaction $^{197}$Au+$^{197}$Au at an
incident beam energy of 400 MeV/nucleon and an impact parameter of
$b=7$ fm.} \label{ek}
\end{figure}
Fig.~\ref{ek-np} shows the effect of the symmetry energy on the
evolution of free n/p ratio. At first stage, values of n/p ratio
with the soft symmetry energies (x= 1) are higher than that with
the stiff symmetry energies (x= 0). This is because at the
beginning of the reaction, compressed nuclear matter densities are
lower than normal nuclear density, the n/p ratio at this stage
just reflects the low-density behavior of the symmetry energy.
After that, nuclear matter's density reaches above normal density
and more free nucleons emit. Finally, the free n/p ratio reflects
high-density behavior of the symmetry energy, thus the values of
free n/p ratio with the stiff symmetry energy (x= 0) are larger
than that with the soft symmetry energy (x= 1). Nevertheless the
discrepancy with the two symmetry energies is quite smaller.

Another common probe is the emission of pion. In this study pions
are created by the decay of the resonances $\Delta(1232)$ and
$N^{*}(1440)$ produced in dense matter since the cross section of
direct pion production is very small at the considered energies.
The reaction channels related to pion production and absorption
are given as follows:
\begin{eqnarray}
&& NN \longrightarrow NN, \nonumber\\ && NR \longrightarrow NR,
\nonumber\\ && NN \longleftrightarrow NR, \nonumber\\ && R
\longleftrightarrow N\pi,
\end{eqnarray}
here $R$ denotes $\Delta $ or $N^{\ast }$ resonances and the
energy and momentum dependent decay width is used in the present
work. Furthermore, the $(\pi^-/\pi^+)_{like}$ ratio is adopted and
based on the dynamics of pion resonance productions and decays, it
read as $ (\pi^-/\pi^+)_{like}\equiv
\frac{\pi^-+\Delta^-+\frac{1}{3}\Delta^0+\frac{2}{3}N^{*0}}
{\pi^++\Delta^{++}+\frac{1}{3}\Delta^++\frac{2}{3}N^{*+}}.$
Fig.~\ref{ek-pi} shows the charged pion ratio as a function of
time with different symmetry energies. It is interesting to see
that at time t $\sim$11 fm/c, there is a cross similar to free n/p
ratio as shown in Fig.~\ref{ek-np}, but for the charged pion ratio
the values are larger for the stiff symmetry energy (x= 0) before
11 fm/c and become smaller after 11 fm/c. Overall, the ratio
$\pi^-/\pi^+$ is only slightly enhanced by the soft symmetry
energy. From Fig.~\ref{ek}, it is seen that very little effects of
the symmetry energy are shown on both the ratio $n/p$ and the
ratio $\pi^-/\pi^+$. This is partially due to the fact that the
density of compressed region in mid-central collisions is lower
than that in central reactions. Thus rather delicate observational
methods have to be found to probe the high density behavior of the
nuclear symmetry energy.

\begin{figure} \centering
\subfigure[~The $n/p$ ratio as a function of the azimuthal angle.]
{ \label{ang-np}
\includegraphics[height=5.5cm,width=0.8\columnwidth]{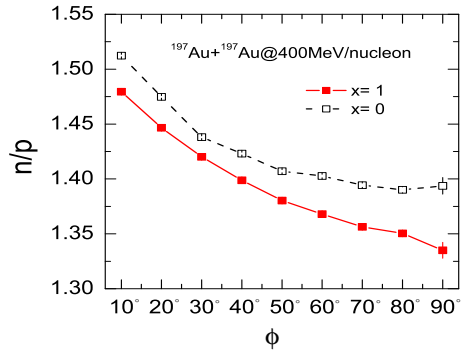}
} \subfigure[~The $\pi^-/\pi^+$ ratio as a function of the
azimuthal angle.] { \label{ang-pi}
\includegraphics[height=5.5cm,width=0.8\columnwidth]{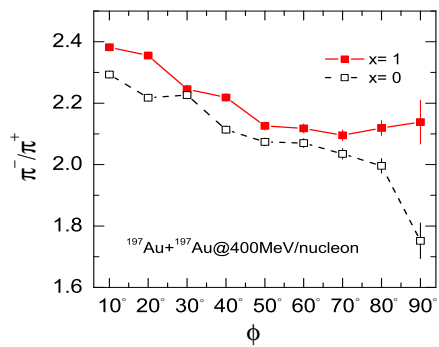}
} \caption{(Color online) The cone-azimuth dependence of the effect of the
symmetry energy on the $n/p$ and $\pi^-/\pi^+$ in the reaction
$^{197}$Au+$^{197}$Au with an impact parameter of $b=7$ fm}
\label{ang}
\end{figure}
Anisotropic distribution of the azimuthal angle of the particle
emission in heavy-ion collisions has been observed in a number of
experiments. The analysis of the angle distributions can help us
to extract the information of the hot and dense nuclear matter.
The azimuth dependence of the effect of the symmetry energy on the
$n/p$ and $\pi^-/\pi^+$ ratios are plotted in Fig.~\ref{ang-np}
and Fig.~\ref{ang-pi}, where $\phi$ is the cone-azimuthal angle of
the emitted particle with respect to the reaction plane and
defined as $\phi= \arcsin(|p_y|/\sqrt{p_x^{2}+p_y^2+p_z^2})$. From
Fig.~\ref{ang}, we can clearly see that with increasing $\phi$ the
ratios both exhibit more sensitivity to the symmetry energy.

\begin{figure} \centering
\subfigure[~Kinetic energy distribution of the ratio $n/p$
perpendicular to the reaction plane ($\phi\sim 90^{0}$).] {
\label{90-np}
\includegraphics[height=5.5cm,width=0.8\columnwidth]{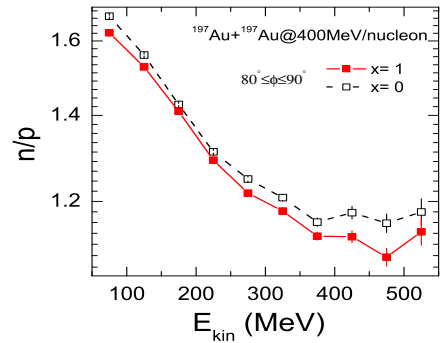}
} \subfigure[~Kinetic energy distribution of the ratio
$\pi^-/\pi^+$ perpendicular to the reaction plane ($\phi\sim
90^{0}$).] { \label{90-pi}
\includegraphics[height=5.5cm,width=0.8\columnwidth]{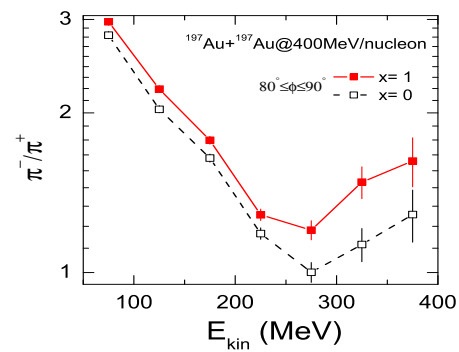}
} \caption{(Color online) Kinetic energy dependence of the ratios of the
particles emitted in the direction perpendicular to the reaction
plane ($\phi\sim 90^{0}$).} \label{90}
\end{figure}
Plotted in Fig.~\ref{90} is the kinetic energy dependence of the
ratios of $n/p$ and $\pi^-/\pi^+$ in the direction perpendicular
to the reaction plane. Compared with that shown in Fig.~\ref{ek},
effects of the symmetry energy on the two observables are indeed
enlarged evidently and increase with particle's kinetic energy,
e.g., in high kinetic region for charged pion ratio shown in
Fig.~\ref{90-pi}, effect of symmetry energy can be as high as more
than 20\%, whereas its effect on the total charged pion ratio
shown in Fig.~\ref{ek} is about 5\%.

This enlargement of the effect of the symmetry energy is
reasonable and consistent with the findings in Ref.~\cite{msup}.
It is well known that the squeeze-out of nuclear matter occurs in
non-central heavy-ion collisions. Particles emitted in the
direction perpendicular to the reaction plane carry direct
information about the high density phase and thus reflect the high
density behavior of nuclear symmetry energy. Moreover, the
energetic particles are mostly produced in the high density
region. As a result the free n/p or $\pi^-/\pi^+$ ratios at high
kinetic energies in the direction perpendicular to the reaction
plane exhibit significant sensitivity to the symmetry energy.

%\section{Summary}

In conclusion, based on  the isospin-dependent
Boltzmann-Uehling-Uhlenbeck transport model, the semi-central
collision of $^{197}$Au+$^{197}$Au at an incident beam energy of
400 MeV/nucleon are studied. It is found that the sensitivities to the symmetry energy of the charged pion ratio, as well as the free
neutron/proton ratio, arise with increase of
cone-azimuthal angle of the emitted particles. And in the
direction perpendicular to the reaction plane, the $\pi^{-}/\pi
^{+}$ ratio or free n/p ratio especially at high kinetic energies
exhibits significant sensitivity to the symmetry energy.

\section*{Acknowledgments}

We acknowledge support by the computing server C3S2 in Huzhou
Teachers College. The work is supported by the National Natural
Science Foundation of China (Grant Nos. 11375239, 11105035,
11375062, 11175219, 11175074), the Zhejiang Provincial Natural
Science Foundation of China (No. Y6110644), the Qian-Jiang Talents
Project of Zhejiang Province (No. 2010R10102), the Scientific
Research Fund of Zhejiang Provincial Education
Department(Y201222917), and the Construct Program of the Key Discipline in Zhejiang Province.

\end{document}